\documentclass[]{elsarticle}

\usepackage{lineno, hyperref}
\modulolinenumbers[5]

\usepackage{amsmath,amssymb,amsthm,amsfonts}

\hypersetup{
    colorlinks,
    linkcolor={red!50!black},
    citecolor={blue!50!black},
    urlcolor={blue!80!black}
}

\usepackage{xcolor}
\newcommand{\ve}[1]{\overrightarrow{#1}}
\newcommand{\matr}[1]{\mathbf{#1}}
\usepackage{bbm}
\newcommand{\mone}{\mathbbm{1}}

\usepackage[ruled,algo2e,linesnumbered,algonl]{algorithm2e}
\usepackage[free-standing-units,group-separator={,}]{siunitx}

\SetCommentSty{mycommfont}

\begin{document}

\begin{frontmatter}

\title{Toward transient finite element simulation of thermal deformation of machine tools in real-time\tnoteref{mytitlenote}}
\tnotetext[mytitlenote]{This work was funded by the German Research Foundation as part of the CRC/TR 96.}

\author[tudresden]{Andreas Naumann\corref{mycorrespondingauthor}}
\cortext[mycorrespondingauthor]{Corresponding author}
\ead{Andreas.Naumann@tu-dresden.de}

\author[unileeds]{Daniel Ruprecht}
\ead{d.ruprecht@leeds.ac.uk}

\author[tudresden]{Joerg Wensch}
\ead{Joerg.Wensch@tu-dresden.de}

\address[tudresden]{Institut for Scientific Computing, Technische Universit\"at Dresden, Dresden, Germany}
\address[unileeds]{School of Mechanical Engineering, University of Leeds, Woodhouse Lane, LS2 9JT, Leeds, United Kingdom}

\begin{abstract}
Finite element models without simplifying assumptions can accurately describe the spatial and temporal distribution of heat in machine tools as well as the resulting deformation.
In principle, this allows to correct for displacements of the Tool Centre Point and enables high precision manufacturing.
However, the computational cost of FEM models and restriction to generic algorithms in commercial tools like ANSYS prevents their operational use since simulations have to run faster than real-time.
For the case where heat diffusion is slow compared to machine movement, we introduce a tailored implicit-explicit multi-rate time stepping method of higher order based on spectral deferred corrections.
Using the open-source FEM library DUNE, we show that fully coupled simulations of the temperature field are possible in real-time for a machine consisting of a stock sliding up and down on rails attached to a stand.
\end{abstract}


\end{frontmatter}

\linenumbers

%
%
\section{Introduction}
Machine tools that are capable of correcting for displacements of the Tool Centre Point (TCP) caused by thermal expansion are a promising approach for high precision manufacturing (other approaches involve, e.g., design modification or thermal-error control)~\cite{LiEtAl2015}.
Most machine tools these days are ``intelligent'' and employ sensors to measure temperature.
Compensating for thermal errors requires knowledge of the the deviation from the machine's reference temperature.

Since the moving parts of a machine result in strongly position and time dependent heat sources and deformations~\cite{Mayr2009,MayrEtAl2012}, this knowledge should ideally include spatial and temporal variations to account for position-dependent heating and transient effects.
Since sensors can only provide data at isolated points, computational models are required to complement measured data and obtain accurate temperature distributions.
Obviously, to allow for the correction of thermal errors during operations, any model to be used for online error compensation has to run faster than real-time in the sense that the ``look-ahead factor'' satisfies
\begin{equation*}
	\eta = \frac{\text{simulated time}}{\text{wall-clock time}} > 1.
\end{equation*}
If,~e.g., we simulate the machine over \SI{10}{\second} and this simulation requires \SI{5}{\second} to run, we achieve a look-ahead factor of $\eta = 2$.
The larger $\eta$, the further into the future the simulation can ``see''.
We focus on the case where the movement of the machine is fast compared to diffusive heat transport and simulated time equals multiple complete machine cycles.

Finite element models (FEM) are derived from first principles and can thus provide a reliable and detailed description of heat transfer and diffusion, even though accurate specification of boundary conditions can be a challenge~\cite{LiEtAl2017}.
Accurate transient finite element models are very useful as they can provide spatially and temporally resolved temperature fields for machines with complex designs and geometries~\cite{MayrEtAl2012,HaitaoEtAl2007}.
In contrast to empirical approaches~\cite{FletcherEtAl2015,MaEtAl2017}, the parameters in FEM are physical quantities that can, at least theoretically, be measured.
Since reduced models are typically machine-specific, their derivation also comes with a high cost in terms of person hours.
In contrast, the mesh for FEM models can be generated automatically, e.g. from CAD files, even for machines with complex geometries. 

The disadvantage of FEM models is their high computational cost, which is why often reduced models are employed, sacrificing accuracy or generality for speed.
Running full time-dependent FEM models is considered too computationally expensive to be possible in real-time: ``application of the original FE-models without any simplifications [...] for model-based control-integrated correction is very time-consuming and thus impractical''~\cite{Galant2016}.
Despite only resolving one machine part and employing a time-averaged heat source instead of a full coupling, Galant et al. report a computation time of around 5 hours to simulate a milling machine with 16,626 degrees-of-freedom over 16 hours using ANSYS (corresponding to $\eta = 3.2$).
To the best of the authors' knowledge, there are no reports of simulations solving in real-time the fully coupled transient FEM problem for a machine with moving parts without simplifications.
Recent review papers also make no mention of such efforts~\cite{LiEtAl2015,MayrEtAl2012,LiEtAl2008}.
A combination of finite differences and FEM, called FDEM, has been proposed that reduces computational effort but still relies on the use of macro elements to reduce the size of the solved system~\cite{Mayr2009}. 
With respect to FDM and FEM, in a review from 2017, Cao et al. state that ``[...], due to the low efficiency, the computational models were rarely used in online thermal error compensation''~\cite{CaoEtAl2017}, mentioning only approaches that rely on steady-state FEM models~\cite{DenkenaEtAl2007,DuEtAl2015}.

A key reason is probably that while widely used commercial proprietary software like ANSYS~\cite{ansys} is easy to use, this simplicity comes with a performance penalty and restriction to generic numerical methods that do not consider the special structure of the problem.
To solve the fully coupled problem in ANSYS, e.g., only implicit Euler is applicable~\cite{Naumann2016}.
While implicit Euler is a robust and widely used time stepping method, it is only first order accurate and does not take into account the different time scales involved, leaving room for substantial efficiency gains by using more tailored algorithms of higher order.

\paragraph{Contributions}
We demonstrate that accurate faster than real-time simulations with a full transient FEM model with \SI{16626}{} degrees of freedom are possible by implementing a tailored higher order multi-rate time stepping method in the open-source finite element library DUNE~\cite{bastian2010infrastructure,engwer2016concepts,SanderDune}. 
While open-source FEM libraries are typically more difficult to use than commercial packages, they are flexible and offer efficient implementations of spatial discretisations and solvers and can be tailored to specific problems.

Our time stepping method is based on multi-rate spectral deferred corrections (MRSDC)~\cite{DuttEtAl2000,BourliouxEtAl2003,LaytonMinion2004,BouzarthMinion2010,EmmettEtAl2014}.
It combines implicit treatment of heat diffusion over larger time steps with explicit integration of the machine movement over smaller steps.
This avoids stability issues from the diffusive term, maintains accuracy for the fast dynamics induced by machine movement and avoids the need to reassemble a Jacobian in each step.
We demonstrate that the new method can substantially improve computational efficiency.
For a look-ahead factor of $\eta = 10$, implicit Euler provides time discretisation errors of about $20\%$ which is probably too inaccurate to compute useful information about the machine deformation.
In contrast, for the same value of $\eta$, MRSDC is about an order of magnitude more accurate, yielding an error of about $3\%$.
For a smaller look-ahead factor of $\eta = 2$, implicit Euler can provide errors of about $1\%$ while MRSDC is again about an order of magnitude more accurate, providing an error below $0.1\%$.

%
%
\section{Description of the problem}
\begin{figure}[!t]
	\centering
 	\includegraphics[width=.6\textwidth]{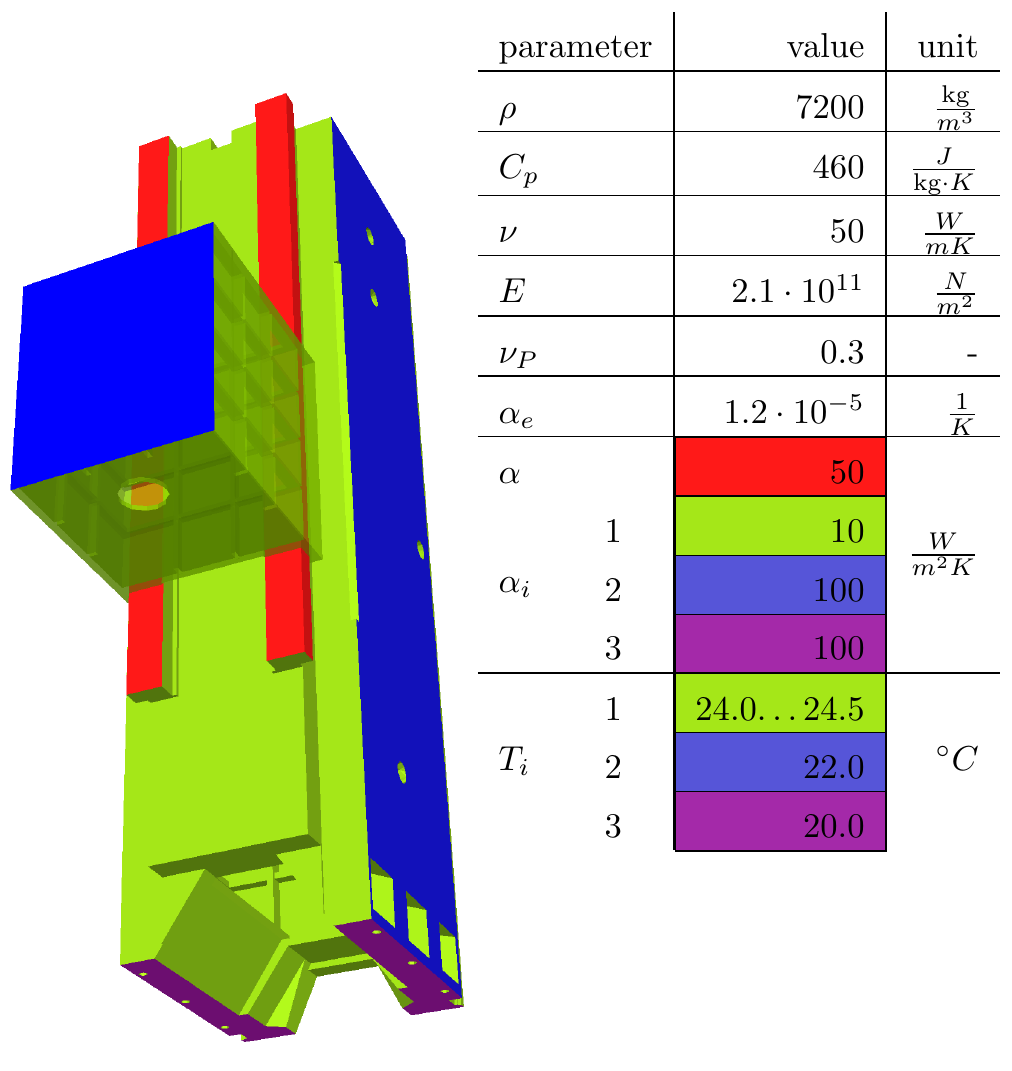}
 	\caption{Fixed stand and stock moving along the rails. Colors indicate different types of boundary conditions: cooling at the right side of the stand and the back of the stock (blue), heat exchange with the floor at the bottom (violet), heat exchange between stand/stock plus heat generated by friction (red) and heat exchange with the environment (green) at all other boundaries.}
 	\label{fig:coloredBoundary}
\end{figure}

Figure \ref{fig:coloredBoundary} shows the configuration of the machine. 
The stand and rail are fixed and the corresponding computational domain is labeled $\Omega_{\text{fix}}$.
The stock moves up and down along the rail (red surface in Figure~\ref{fig:coloredBoundary}) and we refer to this part of the domain as $\Omega_{\text{mov}}$.
The equations modelling diffusion of heat within the two parts read
\begin{subequations}
\label{eq:heat-diffusion}
\begin{align}
	\rho {C_p} \partial_{t} T_{\text{fix}} &=\nu \Delta T_{\text{fix}} \ \text{in} \ \Omega_{\text{fix}} \\
	\rho {C_p} \partial_{t} T_{\text{mov}} &= \nu \Delta T_{\text{mov}}  \ \text{in} \ \Omega_{\text{mov}}(t).
\end{align}
\end{subequations}

Both geometries are coupled through the heat flux at the moving common boundary segment $\Gamma_{\text{R}}(t)=\Gamma_{\text{mov}}(t)\cap \Gamma_{\text{fix}}$ at the rail. 
In Figure~\ref{fig:coloredBoundary}, $\Gamma_{\text{R}}(t)$ corresponds to the part of the top of the rail covered by the stock at time $t$.
At $\Gamma_R$ we have heat exchange between stock and rail and heat generation due to friction.
For the sake of simplicity, we assume thermal isolation at the rest of the rail. 
Put together, we obtain boundary conditions
\begin{subequations}
\begin{align}
	\nu \nabla T_{\text{fix}} \cdot \ve{n} &= 0 \ \text{on} \ \Gamma_{\text{fix,rail}} \setminus \Gamma_{\text{mov}}(t)\\
	\nu \nabla T_{\text{mov}} \cdot \ve{n} &= \alpha \left( T_{\text{fix}} - T_{\text{mov}} \right) + \frac{\eta(t)}{2} \ \text{on} \ \Gamma_{\text{R}}(t) \\
	\nu \nabla T_{\text{fix}} \cdot \ve{n} &= \alpha \left( T_{\text{mov}} - T_{\text{fix}} \right) + \frac{\eta(t)}{2} \ \text{on} \ \Gamma_{\text{R}}(t).
\end{align}
\end{subequations}
We consider here that case where both the moving and fixed part are made of the same material so that $\nu_{\text{fix}} = \nu_{\text{mov}} = \nu$ but using different values for conductivity would be straightforward.

Both domains are also thermally coupled to the surrounding air and a cooling equipment, modelled by Robin boundary conditions at the static pieces of the machine
\begin{subequations}
\begin{align}
\nu\nabla T_{\text{fix}} \cdot \ve{n} &=\alpha_{\text{i}} \left( T_{\text{i}}-T_{\text{fix}}\right)\ \text{on} \ \Gamma_{\text{fix,i}} \\
 \nu \nabla T_{\text{mov}} \cdot \ve{n} &=\alpha_{\text{i}} \left( T_{\text{i}}-T_{\text{mov}}\right) \ \text{on} \ \Gamma_{\text{mov,i}}
\end{align}
\end{subequations}
for $i = 1, 2, 3$.
Here, index $i=1$ represents the boundary where heat is exchanged with the environment (green in Figure~\ref{fig:coloredBoundary}), index $i=2$ the boundary where cooling is applied (blue in Figure~\ref{fig:coloredBoundary}) and finally $i=3$ heat exchange with the floor (violet in Figure~\ref{fig:coloredBoundary}).
Each boundary uses a different value for $T_{\text{i}}$ and $\alpha_{\text{i}}$.
The environmental temperature is assumed to be equal to \SI{24}{\celsius} at the floor with a slight increase of \SI{0.5}{\celsius} over the \SI{2}{\meter} distance to the top of the stand, modelling a sunlit workshop on a warm day.

Both domains are meshed independently and equations~\eqref{eq:heat-diffusion} are discretized using linear finite elements. 
Meshes $\Omega_{\text{fix}}$ and $\Omega_{\text{mov}}$ have different basis and test function spaces $V_{\text{fix}}$ and $V_{\text{mov}}$. 
Multiplying equations~\eqref{eq:heat-diffusion} with the corresponding test functions, integrating each domain separately and inserting the boundary conditions yields
\begin{subequations}
 \begin{align}
  \label{eq:femContFix} \rho {C_p} \int_{\Omega_{\text{fix}}} \partial_t T_{\text{fix}} \varphi_{\text{fix}} dx =& -\nu \int_{\Omega_\text{fix}} \nabla T_{\text{fix}} \nabla \varphi_{\text{fix}} dx \cr
  +&\int_{\Gamma_{\text{fix,env}}} \alpha_{\text{fix}} (T_{\text{env}}-T_{\text{fix}})\varphi_{\text{fix}} dS \cr
   +&\int_{\Gamma_{\text{R}}(t)} \left(\alpha (T_\text{mov}-T_\text{fix})+\frac{\eta(t)}{2}\right)\varphi_{\text{fix}} dS
 \end{align}
 \begin{align}
  \label{eq:femContMov} \rho {C_p} \int_{\Omega_{\text{mov}}} \partial_t T_{\text{mov}} &\varphi_{\text{mov}} dx = -\nu \int_{\Omega_\text{mov}} \nabla T_{\text{mov}} \nabla \varphi_{\text{mov}} dx \cr
  +&\int_{\Gamma_{\text{mov,env}}} \alpha_{\text{mov}} (T_{\text{env}}-T_{\text{mov}})\varphi_{\text{mov}} dS \cr
   +&\int_{\Gamma_{\text{R}}(t)} \left(\alpha (T_\text{fix}-T_\text{mov})+\frac{\eta(t)}{2}\right)\varphi_{\text{mov}} dS
 \end{align}
\end{subequations}
for every test function $\varphi_{\text{fix}}$ and $\varphi_{\text{mov}}$ respectively. 
By representing the solutions $T_{\text{fix}}(x)=\sum_k \ve{T}_{\text{fix},k}(t) \varphi_{\text{fix},k}(x)$ and $T_{\text{mov}}(x)=\sum_k \ve{T}_{\text{mov},k} \varphi_{\text{mov},k}(x)$ in basis functions on the corresponding mesh we can write the continuous equations \eqref{eq:femContFix} and \eqref{eq:femContMov} in their discrete forms
\begin{subequations}
 \label{eq:femDiscr}
 \begin{align}
 \matr{M}_{\text{fix}} \partial_t \ve{T}_{\text{fix}}=& \matr{A}_\text{fix} \ve{T}_\text{fix}+ \matr{B}_{\text{fix,env}} \ve{T}_\text{fix} + \ve{b}_{\text{fix,env}}\cr
  +&\matr{M_B}_{\text{fix}}(t)\ve{T}_{\text{fix}}+\matr{C_B}_\text{mov,fix}(t)\ve{T}_{\text{mov}}+\ve{b}_\text{fix}(t)\\
  \matr{M}_{\text{mov}} \partial_t \ve{T}_{\text{mov}}=& \matr{A}_\text{mov} \ve{T}_\text{mov}+ \matr{B}_{\text{mov,env}} \ve{T}_\text{mov} +\ve{b}_\text{mov,env}\cr +&\matr{M_B}_{\text{mov}}\ve{T}_{\text{mov}}+\matr{C_B}_\text{fix,mov}(t)\ve{T}_{\text{fix}}+\ve{b}_\text{mov}(t).
\end{align}
\end{subequations}
To avoid duplication, we use a generic subscript X instead of ``mov'' and ``fix''  when expressions are identical on both parts.
In both equations we have the standard mass matrix $\matr{M}_X=\int_{\Omega_X} \varphi_X \varphi_X dx$ and the discrete Laplacian ${\matr{A}_X}=-\nu \int_{\Omega_X} \nabla \varphi_X \cdot \nabla \varphi_X dx$ for every test function $\varphi_X$.
We have split the contributions from the environment in two parts. 
The first part 
\begin{equation}
\matr{B}_{X, \text{env}}=-\int_{\Gamma_{X,\text{env}}} \alpha_X \varphi_X \varphi_X dS
\end{equation}
depends on the machine temperature while the second part 
\begin{align} 
  \ve{b}_{\text{X, env}} =&\int_{\Gamma_{\text{X,env}}} \alpha_{\text{X}} T_{\text{env}}\varphi_{\text{X}} dS
\end{align}
does not.
The time dependent source term
 \begin{align}
  \ve{b}_{\text{X}}(t)=&\int_{\Gamma_R(t)} \frac{\eta(t)}{2}\varphi_{\text{X}} dS
 \end{align}
models heat generation through friction~\cite{Naumann2016}.
We split the term $\matr{M}_X(t)\mathbf{T}_X+\matr{C_B}_{Y,X}\ve{T}_Y$ modelling heat exchange between stand and stock in two terms. 
The coefficients of the first part
\begin{align}
 \matr{M}_X=&-\int_{\Gamma_R(t)} \alpha \varphi_X \varphi_X dS
\end{align}
are similiar to the matrix from the Robin boundary condition, whereas the matrix of the second part
\begin{align}
 \label{eq:defCB} \matr{C_B}_{Y,X}=& \int_{\Gamma_R(t)} \alpha \varphi_Y\varphi_X dS
\end{align}
contains basis functions from both domains. 
This is the term which couples the temperature fields of both machine components. 
Since both parts are meshed independently, the meshes do not match at the interface. 
Therefore, we have to compute the intersections of both meshes using existing methods for grid coupling in DUNE~\cite{bastian2010infrastructure, engwer2016concepts} to evaluate the boundary integrals in~\eqref{eq:defCB}. 

Now, we combine both equations in~\eqref{eq:femDiscr} into one coupled system 
\begin{align}
\label{eq:coupledFEM} \left(\begin{matrix}
  \matr{M}_{\text{fix}} & 0 \cr 0 & \matr{M}_{\text{mov}}
 \end{matrix} \right)\partial_t \ve{T} =&\left(\begin{matrix} \matr{A}_{\text{fix}}+\matr{B}_{\text{fix,env}} & 0 \cr 
 0 & \matr{A}_{\text{mov}}+\matr{B}_{\text{mov,env}}\end{matrix}\right)\ve{T} + 
  \left(\begin{array}{c}
    \ve{b}_{\text{fix,env}} \cr
    \ve{b}_{\text{mov,env}}
  \end{array}\right) \cr
  +&\left(\begin{matrix} \matr{M_B}_{\text{fix}}(t) & \matr{C_B}_{\text{mov,fix}}(t) \cr 
 \matr{C_B}_{\text{mov,fix}}^T(t) & \matr{M_B}_{\text{mov}}\end{matrix}\right)\ve{T} + 
  \left(\begin{array}{c}
    \ve{b}_{\text{fix}}(t) \cr
    \ve{b}_{\text{mov}}(t)
  \end{array}\right)
\end{align}
by introducing $\ve{T}=\left( \begin{array}{c} \ve{T}_{\text{fix}} \cr \ve{T}_{\text{mov}} \end{array}\right)$. 
We also replace the mass matrices $\matr{M}_\text{fix}$ and $\matr{M}_\text{mov}$ by their row sum-lumped version \cite{Zienkiewicz2005}.

In preparation for the introduction of the multi-rate time stepping in the next section, we split the right hand side function into the following parts
\begin{align}
 \ve{f}^I(\ve{T})=&\left(\begin{matrix} \matr{A}_{\text{fix}}+\matr{B}_{\text{fix,env}} & 0 \cr 
 0 & \matr{A}_{\text{mov}}+\matr{B}_{\text{mov,env}}\end{matrix}\right)\ve{T} \\
 \ve{g}(t)=&\left(\begin{array}{c}
    \ve{b}_{\text{fix,env}} \cr
    \ve{b}_{\text{mov,env}}\end{array}\right) \\
 \ve{f}^E(\ve{T},t)=&\left(\begin{matrix} \matr{M_B}_{\text{fix}}(t) & \matr{C_B}_{\text{mov,fix}}(t) \cr 
 \matr{C_B}_{\text{mov,fix}}^T(t) & \matr{M_B}_{\text{mov}}\end{matrix}\right)\ve{T} + 
  \left(\begin{array}{c}
    \ve{b}_{\text{fix}}(t) \cr
    \ve{b}_{\text{mov}}(t)
  \end{array}\right).
\end{align}
With this notation~\eqref{eq:coupledFEM} can compactly be written as
\begin{equation}
	\label{eq:ivp}
	\matr{M} \partial_t \ve{T}(t) = \ve{F}(\ve{T}(t), t) = \ve{f}^{I}(\ve{T}(t)) + \ve{f}^E(\ve{T}(t), t) + \ve{g}(t).
\end{equation}
Since we consider the regime where heat diffusion is slow compared to machine movement, $\ve{f}^I$ represents a slow process.
In contrast, $\ve{f}^E$ represents the fast coupling process and generation of heat from friction.
Evaluating $\ve{f}^E$ requires detecting the intersection of finite elements at the interface between the two components, which can be expensive.
Lastly, $\ve{g}(t)$ models heat exchanges with the environment, floor and cooling, which are also slow relative to the movement of the machine.

%
%

%
%
\section{Numerical time stepping method}
\label{seq:numericalMethod}
In this section, we present a time stepping algorithm with a problem-specific multi-rate splitting based on spectral deferred corrections~\cite{DuttEtAl2000} that will reduce solution times significantly compared to a standard implicit Euler method. 

The term $\ve{f}^I$ models diffusion of heat.
Resolving it accurately requires a time step $\Delta t = \mathcal{O}(\Delta x)$ while stability for an explicit integrator requires $\Delta t = \mathcal{O}(\Delta x^2)$.
To resolve all geometrical features of the machine the mesh has many small elements with diameters of the order of \SI{2e-4}{\meter} whereas the stand is \SI{2}{\meter} high and has a \SI{0.5}{\meter} by \SI{0.5}{\meter} base.
Given the values for $\nu$, $\rho$, $C_p$ in Figure~\ref{fig:coloredBoundary}, an explicit integrator would require a time step
\begin{equation}
	\Delta t \leq \frac{\rho C_p \Delta x^2}{\nu} = \SI{0.0027}{\second}
\end{equation}
for stability, which is orders of magnitude too small to be efficient.
Therefore, $f^I$ is treated implicitly with a larger time step.
The term $\ve{g}$ is independent of $\ve{T}$ and models heat exchange with the environment which is a slow process.
Therefore, we use the same large time step as for $\ve{f}^I$.

In contrast, for $\ve{f}^E$, modelling the movement of parts of the machine, we require that $\Delta t = \mathcal{O}(\frac{\Delta x}{v})$ ($v$ being the speed of the machine).
Otherwise, the stock moves across multiple mesh cells in one time step, creating a ``stroboscope effect'' and highly unrealistic temperature distributions~\cite{partzsch_beitelschmidt_2014}.
There is thus no benefit integrating $\ve{f}^E$ implicitly because taking a large time step is impossible anyway.
Furthermore, implicit treatment of this term leads to a time-dependent Jacobian and a potentially large number of evaluations, each of which would require detecting intersections.
To avoid both issues, we integrate $\ve{f}^E$ explicitly but with a smaller time step.

Finally, to achieve better computational efficiency, we want our time stepping method to be at least second order accurate.
Derivation of both implicit-explicit and multi-rate method of higher order is challenging and we employ the spectral deferred corrections framework for this purpose.

\subsection*{Single-rate spectral deferred correction}
Before discussing the multi-rate SDC algorithm, we first describe its single-rate variant.
Consider the initial value problem~\eqref{eq:ivp} over one time step $[t_n, t_{n+1}]$.
Let 
\begin{equation*}
	t_n \leq \tau_1 < \ldots < \tau_M \leq t_{n+1}
\end{equation*}
denote a set of quadrature nodes within the time step.
We denote the distances between nodes by $\Delta \tau_m = \tau_{m} - \tau_{m-1}$ for $m=2, \ldots, M$ and $\Delta \tau_1 = \tau_1 - t_n$.
The analytical solution of~\eqref{eq:ivp} satisfies the integral equations
\begin{equation}
	\label{eq:collocation_continuous}
	\matr{M} \ve{T}(\tau_{m}) = \matr{M} \ve{T}(t_n) + \int_{t_n}^{\tau_m} \ve{F}(\ve{T}(s), s)~ds
\end{equation}
for $m=1,\ldots,M$.
We approximate the integral using a quadrature rule, resulting in the discrete approximations 
\begin{equation}
	\label{eq:collocation}
	\matr{M} \ve{T}_m = \matr{M} \ve{T}(t_n) + \sum_{j=1}^{M} q_{m,j} \ve{F}(\ve{T}_j, \tau_j)
\end{equation}
of~\eqref{eq:collocation_continuous} with $\ve{T}_m \approx \ve{T}(\tau_m)$.
The quadrature weights $q_{m,j}$ are given as integrals over Lagrange polynomials~\cite{RuprechtSpeck2016}.
This approach is known as collocation and the unknowns $\ve{T}_m$ correspond to the stages of a fully implicit Runge-Kutta method with Butcher tableau~\cite[Theorem 7.7]{HairerEtAl1993_nonstiff}.
Theoretically, these can be computed using a Newton-Raphson method to solve the $M$ coupled nonlinear equations~\eqref{eq:collocation} but the large size of the nonlinear system makes this approach impractical for systems with a large number of degrees-of-freedom, in particular semi-discrete partial differential equations.

Instead, spectral deferred corrections employ an iterative procedure which avoids assembly of the full system. 
Each iteration can be computed by a ``sweep'' through the quadrature nodes with a low order method.
Semi-implicit SDC (SISDC)~\cite{Minion2003} starts with an initial prediction step using IMEX-Euler to generate approximate values $\ve{T}^0_m$ from
\begin{align}
	\matr{M} \ve{T}^{0}_{m} &= \matr{M} \ve{T}^{0}_{m-1} + \Delta \tau_m \left( \ve{f}^{I}(\ve{T}^{0}_{m}) + \ve{g}(\tau_{m}) \right) + \Delta \tau_m  \ve{f}^E(\ve{T}^{0}_{m-1}, \tau_{m-1})
\end{align}
for $m=1, \ldots, M$ with $\ve{T}^0_0 = \ve{T}(t_n)$.
This provides a first order accurate approximation of $\ve{T}$ at the quadrature nodes.
Then, to increase the order, SISDC proceeds with the following iterative correction
\begin{align}
	\label{eq:sdc_sweep}
	\matr{M} \ve{T}^{k+1}_{m} &= \matr{M} \ve{T}^{k+1}_{m-1} + \Delta \tau_m \left( \ve{f}^{I}(\ve{T}^{k+1}_{m}) - \ve{f}^{I}(\ve{T}^k_m) \right) \\
			   & + \Delta \tau_m \left( \ve{f}^E(\ve{T}^{k+1}_{m-1}, \tau_{m-1}) - \ve{f}^E(\ve{T}^k_{m-1}, \tau_{m-1}) \right) + I_{m-1}^{m}.
\end{align}
with
\begin{equation}
	I_{m-1}^m := \sum_{j=1}^{M} s_{m,j} \ve{F}(\ve{T}^k_j, \tau_j) \approx \int_{\tau_{m-1}}^{\tau_m} \ve{F}(\ve{T}(s),s)~ds.
\end{equation}
The weights are given by $s_{m,j} := q_{m,j} - q_{m-1,j}$ for $m=2, \ldots, M$ and $s_{1,j} := q_{1,j}$.
Note that since the source term $\ve{g}$ does not change with $k$, $\ve{g}(\tau_{m}) - \ve{g}(\tau_m)$ cancels out but it is considered in the correction steps through $\ve{F}(\ve{T}^k_j, \tau_j)$ in the quadrature term $I_{m-1}^{m}$.

For $k \to \infty$, if the iteration converges and $\ve{T}^{k+1}_m - \ve{T}^k_m \to 0$ for $m=1,\ldots,M$, Equation~\eqref{eq:sdc_sweep} reduces to
\begin{equation}
	\matr{M} \ve{T}_m = \matr{M} \ve{T}_{m-1} + I_{m-1}^{m}.
\end{equation}
Applying this equation recursively shows that the $\ve{T}^{k+1}_m$ converge to the solutions $\ve{T}_m$ of~\eqref{eq:collocation}.
However, the value of SDC stems from the fact that it is not necessary to fully solve the collocation problem. 
It can be shown~\cite{RuprechtSpeck2016} that, if the time step is small enough, each iteration reduces the residual
\begin{equation}
	\label{eq:residual}
	r^k := \max_{m=1,\ldots,M} \left\| \matr{M} \ve{T}_m^k  - \matr{M} \ve{T}(t_{n}) - \sum_{j=1}^{m} I_{m-1}^{m} \right\|
\end{equation}
by a factor of proportional to $\Delta t$.
Therefore, each iteration increases the formal order of the method by one, up to the order of the underlying quadrature rule which depends on $M$ and the chosen type of nodes.
Thus, by adjusting the runtime parameter $K$ and $M$, SISDC allows to generate a split scheme of arbitrary order.

%
%
\subsection*{Multi-rate spectral deferred correction}
\label{seq:MRSDC}
\begin{figure}
	\centering	
	\includegraphics[width=\textwidth]{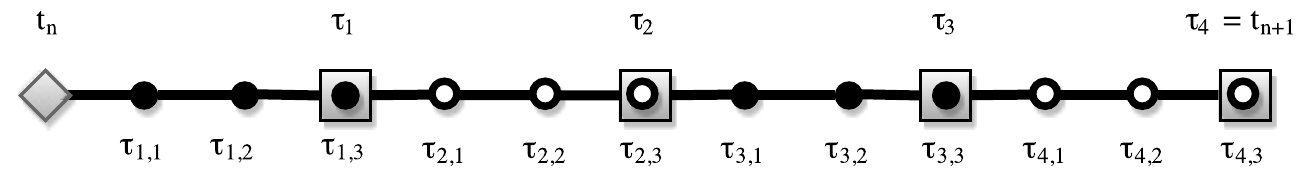}
	\caption{Standard quadrature nodes $\tau_m$, $m=1, \ldots, M$ (grey squares) and embedded quadrature nodes $\tau_{m, p}$, $p=1,\ldots,P$ (black and white circles) within a time step $[t_n, t_{n+1}]$ in multi-rate spectral deferred correction (MRSDC) for $M=4$ standard nodes and $P=3$ embedded nodes. We use no-left equidistant nodes, that is $t_n$ is not a standard quadrature nodes and $\tau_1 = \tau_{1,3}$ is part of the  first set of embedded nodes $(\tau_{1,j})_{j=1, \ldots P}$ in $[t_n, \tau_1]$, but not an embedded node in $[\tau_1, \tau_2]$.}
	\label{fig:mrsdc}
\end{figure}

Multi-rate SDC (MRSDC) has been first introduced by Bourlioux, Layton and Minion~\cite{BourliouxEtAl2003}.
In MRSDC, a set of \emph{embedded} quadrature nodes $\tau_{m, p}$, $p = 1, \ldots, P$, is introduced in between each pair $[\tau_{m-1}, \tau_m]$ of \emph{standard} quadrature nodes as illustrated in Figure~\ref{fig:mrsdc}.
Therefore, we have a total of $M \times P$ nodes
\begin{equation}
	t_n \leq \tau_{1,1} < \ldots < \tau_{1,P} \leq \tau_1 < \ldots \ldots < \tau_{M,P} \leq \tau_M \leq t_{n+1}.
\end{equation}
For simplicity, we assume here that the rightmost quadrature node always coincides with the endpoint of the interval so that $\tau_{M} = t_{n+1}$ and $\tau_{m, P} = \tau_{m}$.
Furthermore, we use equidistant quadrature nodes where $t_n$ is not a standard node (that is, $t_n < \tau_1$) and $\tau_{m-1}$ is not a quadrature node for the embedded nodes $\tau_{m,p}$ (that is, $\tau_{m-1} < \tau_{m,1})$.
While equidistant nodes limit the formal order of the quadrature rule to the number of nodes (instead of, e.g., twice the number of nodes for Gauss-Legendre quadrature), it significantly improves SDC's convergence in the very stiff limit~\cite{Weiser2014}.
The fast changing term $\ve{f}^E$ is approximated by a cumulative sum of the embedded nodes (that is, with small steps) while the slowly changing terms $\ve{f}^I$ and $\ve{g}$ are approximated only at the standard nodes.

Subtracting equations~\eqref{eq:collocation_continuous} for $m$ and $m-1$ yields the ``node-to-node'' variants of the integral equations
\begin{align}
	\label{eq:node-to-node}
	\matr{M} \ve{T}(\tau_m) = \matr{M} \ve{T}(\tau_{m-1}) + \int_{\tau_{m-1}}^{\tau_m} \ve{F}(\ve{T}(s), s)~ds.
\end{align}
Similarly, the integral equations at the embedded nodes read
\begin{equation}
	\label{eq:node-to-node-embedded}
	\matr{M} \ve{T}(\tau_{m,p}) = \matr{M} \ve{T}(\tau_{m,p-1}) + \int_{\tau_{m,p-1}}^{\tau_{m,p}} \ve{F}(\ve{T}(s),s)~ds.
\end{equation}
In addition to the approximations at the standard nodes $\ve{T}_j \approx \ve{T}(\tau_j)$ as in single-rate SDC we now also consider  approximations $\ve{T}_{m,p} \approx \ve{T}(\tau_{m,p})$ of the solution $\ve{T}$ at the embedded nodes.
Then, we approximate the integrals with the following quadrature rules
\begin{equation}
	\label{eq:standard_quad}
	 I_{m-1}^{m} \approx \int_{\tau_{m-1}}^{\tau_m} \ve{F}(\ve{T}(s),s)~ds
\end{equation}
with
\begin{equation}
 I_{m-1}^{m} := \sum_{j=1}^{M} s_{m,j} \left( f^I(\ve{T}_j) + g(\tau_{j}) \right) + \sum_{p=1}^P \hat{s}_{m,p}  f^E(\ve{T}_{m,p}, \tau_{m,p}) 
\end{equation}	
and
\begin{equation}
	\label{eq:embedded_quad}
	 I_{m,p-1}^{p} \approx \int_{\tau_{m,p-1}}^{\tau_{m,p}} \ve{F}(\ve{T}(s),s)~ds
\end{equation}
with
\begin{equation}
	I_{m,p-1}^{p} := \sum_{j=1}^{M} \tilde{s}_{m,p,j} \left( f^I(\ve{T}_j) + g(\tau_j) \right) + \sum_{q=1}^{P} s_{m,p,q} f^E(\ve{T}_{m,q}, \tau_{m,q} ).
\end{equation}
The quadrature weights are defined as follows: let $l_m(s)$ denote the Lagrange polynomials with respect to the standard nodes and $l_{m,p}(s)$ the Lagrange polynomials with respect to one set of embedded nodes, that is
\begin{equation}
	l_m(\tau_j) = \delta_{mj}, \quad m,j = 1, \ldots, M
\end{equation}
and
\begin{equation}
	l_{m,p}(\tau_{m,q}) = \delta_{pq}, \quad m=1, \ldots, M, \ p,q=1,\ldots,P
\end{equation}
with $\delta$ being the Kronecker Delta.
Then, the weights are defined as
\begin{subequations}
\begin{align}
	s_{m,j} &:= \int_{\tau_{m-1}}^{\tau_m} l_j(s)~ds \\
	\hat{s}_{m,p} &:= \int_{\tau_{m-1}}^{\tau_m} l_{m,p}(s)~ds \\
	\tilde{s}_{m,p,j} &:= \int_{\tau_{m,p-1}}^{\tau_{m,p}} l_{j}(s)~ds \\
	s_{m,p,q} &:= \int_{\tau_{m,p-1}}^{\tau_{m,p}} l_{m,q}(s)~ds.
\end{align}
\end{subequations}
Thus, the weights $s_{m,j}$ and $\hat{s}_{m,p,j}$ approximate integrals between standard nodes while $\tilde{s}_{m,p}$ and $s_{m,p,q}$ approximate integrals between embedded nodes, see Table~\ref{tab:weights}.
\begin{table}[t]
\centering
\begin{tabular}{|cc|cc|} \hline
\multicolumn{2}{|c|}{Integral boundaries} & \multicolumn{2}{c|}{Position of function values} \\ \hline \hline
&          & Standard        & Embedded            \\ \hline
& Standard & $s_{m,j}$       &  $\hat{s}_{m,q}$\\
& Embedded & $\tilde{s}_{m,p,j}$ & $s_{m,p,q}$ \\ \hline
\end{tabular}
\caption{Quadrature weights for integrals between standard or embedded nodes depending on whether approximate function values are given at standard or embedded nodes.}
\label{tab:weights}
\end{table}

Now we approximate the continuous integral equations~\eqref{eq:node-to-node} and~\eqref{eq:node-to-node-embedded} with their discrete counterparts
\begin{equation}
	\label{eq:discrete-node-to-node}
	\matr{M} \ve{T}_m =\matr{M} \ve{T}_{m-1} + I_{m-1}^{m}
\end{equation}
and
\begin{equation}
	\label{eq:discrete-node-to-node-embedded}
	\matr{M} \ve{T}_{m,p} = \matr{M} \ve{T}_{m,p-1} + I_{m,p-1}^{p}.
\end{equation}
Note that the integral approximations are consistent in the sense that
\begin{equation}
	I_{m-1}^{m} = \sum_{p=1}^{P} I_{m,p-1}^{p}
\end{equation}
because
\begin{equation}
	s_{m,j} = \sum_{p=1}^{P} \tilde{s}_{m,p,j} \quad \text{and} \quad \hat{s}_{m,p} = \sum_{q=1}^{P} s_{m,p,q}.
\end{equation}

Just as for the single-rate case, we consider the residual~\eqref{eq:residual} at the standard nodes.
In theory, we could solve the $(M+1) \times P$ nonlinear equations~\eqref{eq:discrete-node-to-node} and~\eqref{eq:discrete-node-to-node-embedded} directly for the $\ve{T}_m$ and $\ve{T}_{m,p}$.
However, solving such a large system is impractical so we again rely on an iterative approximation.

\begin{algorithm2e}[t!]
	\caption{Multi-rate SDC prediction step.}\label{alg:mrsdc_pred}
        \SetKwInOut{Input}{input}
         \SetKwInOut{Output}{output}
         \Input{$\ve{T}(t_n)$}
         \Output{$\ve{T}^0_m$ and $\ve{T}^0_{m,p}$ for $m=1,\ldots,M$ and $P=1,\ldots,P$.}
         $\ve{T}^0_0 \leftarrow \ve{T}(t_n)$\\
         \For{$m=1, M$}{
         	\tcc{Implicit step over $[\tau_{m-1}, \tau_m]$.}
         	Solve $\matr{M} \ve{T}^*_m = \matr{M} \ve{T}^0_{m-1} + \Delta t_m  \left( f^I(\ve{T}^*_m) + g(\tau_m) \right)$  \\
			$f^*_m \leftarrow f^I(\ve{T}^*_m) + g(\tau_m)$ \\
			\tcc{Set starting value at $\tau_{m-1}$.}
			$\ve{T}^0_{m,0} \leftarrow \ve{T}^0_{m-1}$\\
			\tcc{Sweep through embedded nodes $\tau_{m,p}$ with explicit Euler.}
			\For{$p=1,P$}{
				$\matr{M} \ve{T}^0_{m,p} = \matr{M} \ve{T}^0_{m,p-1} + \Delta t_{m,p} \left( f^*_m+ f^E(\ve{T}^0_{m,p-1}, \tau_{m,p-1})  \right) $ \\
			}
		\tcc{Update value at $\tau_m$ by overwriting with final value from embedded sweep (since $\tau_{m,P} = \tau_m$).}
		$\ve{T}^0_m \leftarrow \ve{T}^0_{m,P}$ \\
         }
\end{algorithm2e}
We start by generating a first order accurate approximation at all nodes (standard and embedded) by computing the predictor step shown in Algorithm~\ref{alg:mrsdc_pred}.
For every standard step $[\tau_{m-1}, \tau_m]$, we first compute one large implicit Euler step to generate an estimated final value $\ve{T}^*_m$ and compute $f^*_m = f^I(\ve{T}^*_m)$. 
Then, we compute a series of small steps using explicit Euler in $\ve{f}^E$, going from $\tau_{m-1} = \tau_{m, 0}$ to $\tau_{m,P} = \tau_m$.
The implicit term remains fixed to $f^*_m$ throughout.
Finally, we use the final results of the series of small steps $\ve{T}^0_{m,P}$ as $\ve{T}^0_m$, that is as initial value for the next embedded step.
This then provides the initial value $\ve{T}^0_m$ for the next interval from $\tau_m$ to $\tau_{m+1}$ where we start the procedure again with a large implicit step.
As we show later, the predictor step provides a first order accurate approximation.

\begin{algorithm2e}[t!]
	\caption{Multi-rate SDC correction sweep.}\label{alg:mrsdc_sweep}
        \SetKwInOut{Input}{input}
         \SetKwInOut{Output}{output}
         \Input{$\ve{T}(t_n)$ and $\ve{T}^k_m$, $\ve{T}^k_{m,p}$ for $m=1,\ldots,M$, $p=1,\ldots,P$.}
         \Output{updated values $\ve{T}^{k+1}_m$, $\ve{T}^{k+1}_{m,p}$}
         \tcc{Update the integral terms}
         Update $I_{m-1}^{m} $, $m=1, \ldots, M$ according to~\eqref{eq:standard_quad}\\
         Update $I_{m-1,p-1}^{p} $, $m=1,\ldots,M$; $p=1,\ldots,P$ according to~\eqref{eq:embedded_quad}\\
         \tcc{Value at beginning of time step $\tau_0 = t_n$ is brought forward from previous step and remains the same for all iterations $k$.}
         $\ve{T}^{k+1}_0 \leftarrow \ve{T}(t_n)$ \\
         \For{$m=1,M$}{
         	\tcc{Implicit correction step over $[\tau_{m-1}, \tau_m]$. Note that the $g(\tau_m)$ term cancels out but is included in $I_{m-1}^m$.}
         	Solve $\matr{M} \ve{T}^*_m = M \ve{T}^{k+1}_{m-1} + \Delta t_m \left( f^I(\ve{T}^{*}_m) - f^I(\ve{T}^k_{m}) \right) + I_{m-1}^{m}$ \\
		$f^*_m \leftarrow f^I(\ve{T}^*_m)-f^I(\ve{T}^k_m)$\\
		\tcc{Set starting value at $\tau_{m-1}$.}
		$\ve{T}^{k+1}_{m,0} \leftarrow \ve{T}^{k+1}_{m-1}$\\
		\For{$p=1,P$}{
			\tcc{Sweep through embedded nodes $\tau_{m,p}$ with explicit Euler.}
			$ 
			 \begin{aligned}
			\matr{M} \ve{T}_{m,p}^{k+1} = & \matr{M} \ve{T}_{m,p-1}^{k+1} + \Delta t_{m,p} f^*_m \\ &+ \Delta t_{m,p} \left( f^E(\ve{T}^{k+1}_{m,p-1}, \tau_{m,p-1}) - f^E(\ve{T}^k_{m,p-1}, \tau_{m,p-1}) \right) \\ & + I_{m,p-1}^{p}
			\end{aligned}$\\
		}
		\tcc{Update value at $\tau_m$ by overwriting with final value from embedded sweep (since $\tau_{m,P} = \tau_m$).}		
		$\ve{T}^{k+1}_m \leftarrow \ve{T}^{k+1}_{m,P}$\\
         }
\end{algorithm2e}
The order is then increased using the iteration shown in Algorithm~\ref{alg:mrsdc_sweep}.
It proceeds similarly to the predictor step by combining a single large implicit step in $\ve{f}^I$ over $[\tau_{m-1}, \tau_m]$ with $P$ many small explicit steps for the embedded nodes.
Through numerical examples, we will demonstrate the following properties.
\begin{enumerate}
	\itemsep0em
 	\item[(i)] \textbf{Convergence to collocation solution:} the residual~\eqref{eq:residual} decreases geometrically proportional to $\Delta t$ and approximately at the same rate as for single-rate SDC.
	\item[(ii)] \textbf{Order of accuracy:} each iteration increases the formal order by one, up to the order of the approximations of the integral $\min\left\{ M, P \right\}$.
	\item[(iii)] \textbf{Computational efficiency:} multi-rate SDC reduces solution times while maintaining the same accuracy as single-rate SDC or implicit Euler.
	\item[(iv)] \textbf{Smooth temperature profiles:} the smaller time step for the coupling in MRSDC leads to a smoother temperature profile than implicit Euler, which results in more realistic deformations, since those depend on the temperature gradient.
\end{enumerate}
Properties (i) and (ii) are demonstrated for a two dimensional problem of reduced complexity while (iii) and (iv) are demonstrated for the fully coupled 3D machine. 

%
%
\section{Convergence to collocation solution and formal order of accuracy}
\begin{figure}
\begin{minipage}[b]{.49\textwidth}
\includegraphics[width=\textwidth]{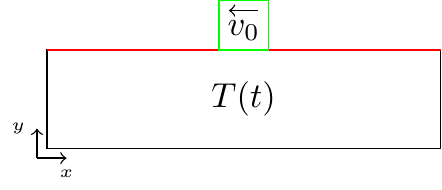}
\caption{The 2d domain with the red boundary line at the top. This red line represents the moving flux with reference temperature $v_0$ and the remaining black lines correspond to zero flux.}
\label{fig:2dProblem_bdCondition}
\end{minipage}\hspace{.5em}
\begin{minipage}[b]{.49\textwidth}
\includegraphics[width=\textwidth]{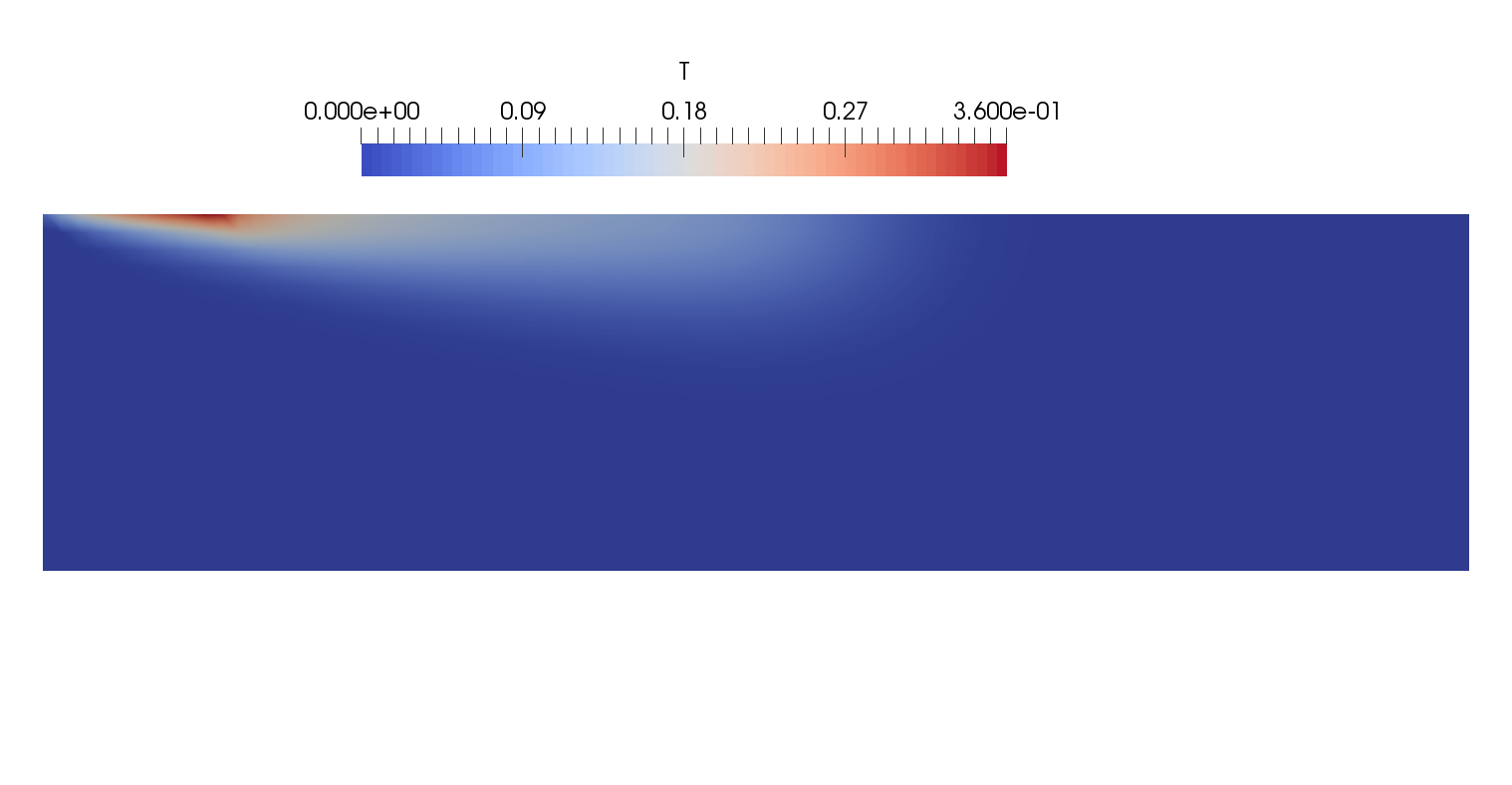}
\vspace{-2.5em}\caption{The temperature field after 20 seconds, i.e. the source square reached the left corner. The temperature tail from the center to the left boundary is clearly visible\vspace{1.5em}}
\label{fig:2dProblem_solution} 
\end{minipage}

\end{figure}

We demonstrate the theoretical properties (i) and (ii) of the method for a simplified 2D version of the full problem that is cheap to solve and allows to easily run simulations for a wide range of  parameters.
The configuration is sketched in Figure~\ref{fig:2dProblem_bdCondition}. 
In this scenario, the stand is a rectangle with variable temperature while the stock is represented as a smaller rectangle of constant temperature $T_0$, gliding left and right.
We neglect the rails and prescribe the heat flux $\nu\nabla T_{\text{fix}}=\alpha(T_0-T)$ at the intersection. 
The stock moves horizontally with velocity $v=-0.1$\SI{}{\meter\per\second}. 
At the remaining boundaries, we assume thermal isolation and apply a zero flux condition.
The stand is discretized with bilinear finite elements. 
In the similar way as for to the 3D problem we obtain the equation
\begin{align}
 \label{eq:2dFEM} \matr{M}\partial_t\ve{T}=&\left(\nu \matr{A}+\alpha \matr{M_B}(t)\right)\ve{T}+\alpha \matr{C_B}(t)\mone T_0\,.
\end{align}
The matrices $\matr{A},\ \matr{M},\ \matr{M_B}$ and $\matr{C_B}$ are analogous to the full problem, but we dropped the subscript ``fix''. 
Our splitting is now straightforward. 
The slow implicit part $f^I$ and the fast part $f^E$ are
\begin{subequations}
\begin{align}
 \ve{f}^I(\ve{T})=&\nu \matr{A}\ve{T} \\
 \ve{f}^E(\ve{T},t)=&\alpha \matr{M_B}(t)\ve{T}+\alpha \matr{C_B}(t)\mone T_0
\end{align}
\end{subequations}
whereas $g(t)=0$ due to the fact the we neglect the thermal exchange with the environment. 
Figure~\ref{fig:2dProblem_solution} shows the temperature field after $t=20$\SI{}{\second} when the stock is located at the left side.
Following the movement of the stock, the temperature increases at the boundary from the center to the left corner with heat dissipating slowly into the stand. 

\begin{figure}
\begin{minipage}[t]{.5\textwidth}
 \includegraphics[width=1.05\textwidth]{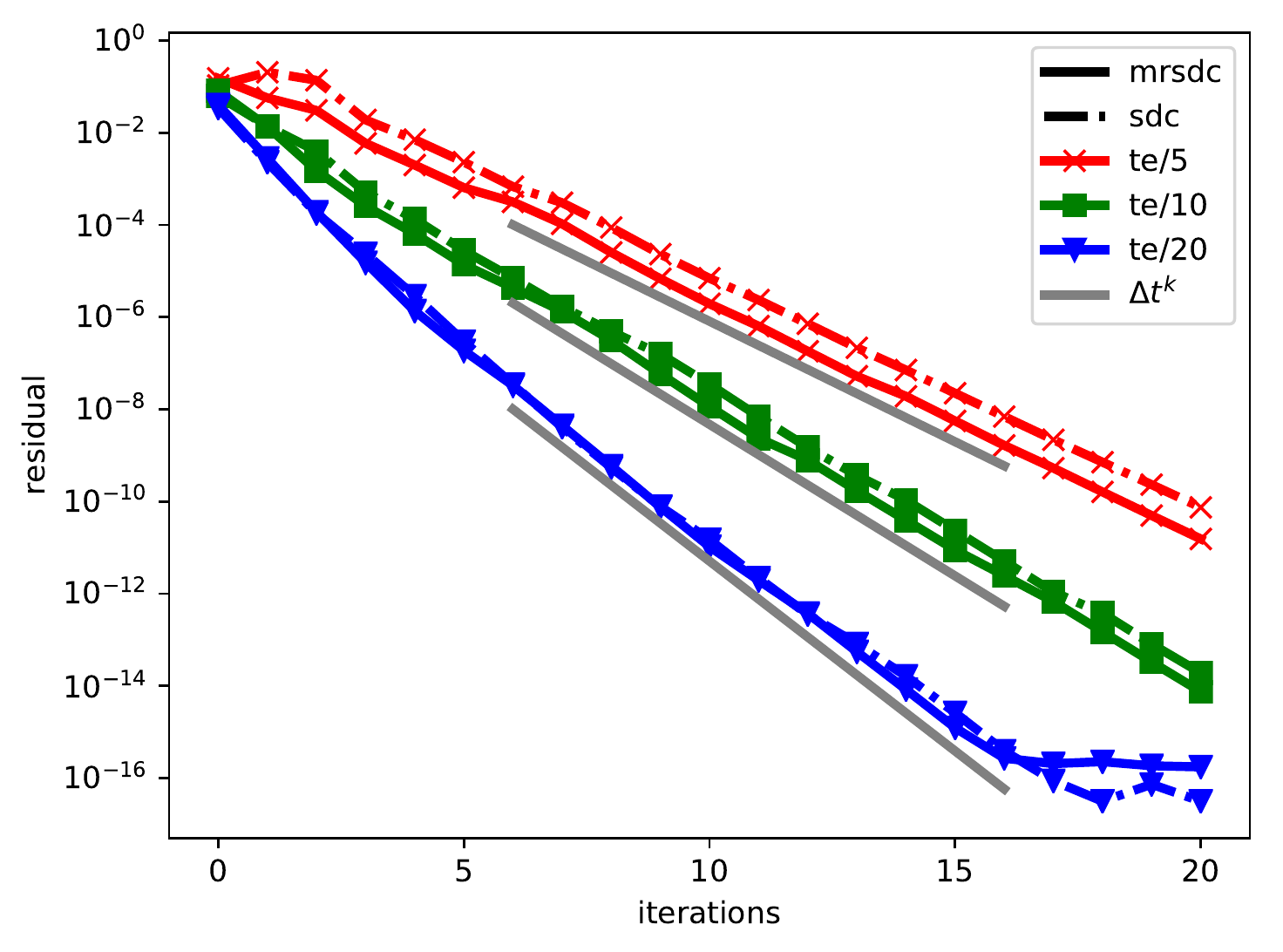}
 \caption{Residual versus iteration count $k$ for SDC with $M=5$ nodes (dashed lines) and MRSDC with $(M,P) = (5,8)$ nodes (solid lines). Both convergence to the collocation solution at approximately the same rate proportional to $\Delta t^k$ (indicated by grey lines).}
  \label{fig:residual_test}
\end{minipage} 
\hspace{0.5em}
\begin{minipage}[t]{.5\textwidth}
 \includegraphics[width=\textwidth]{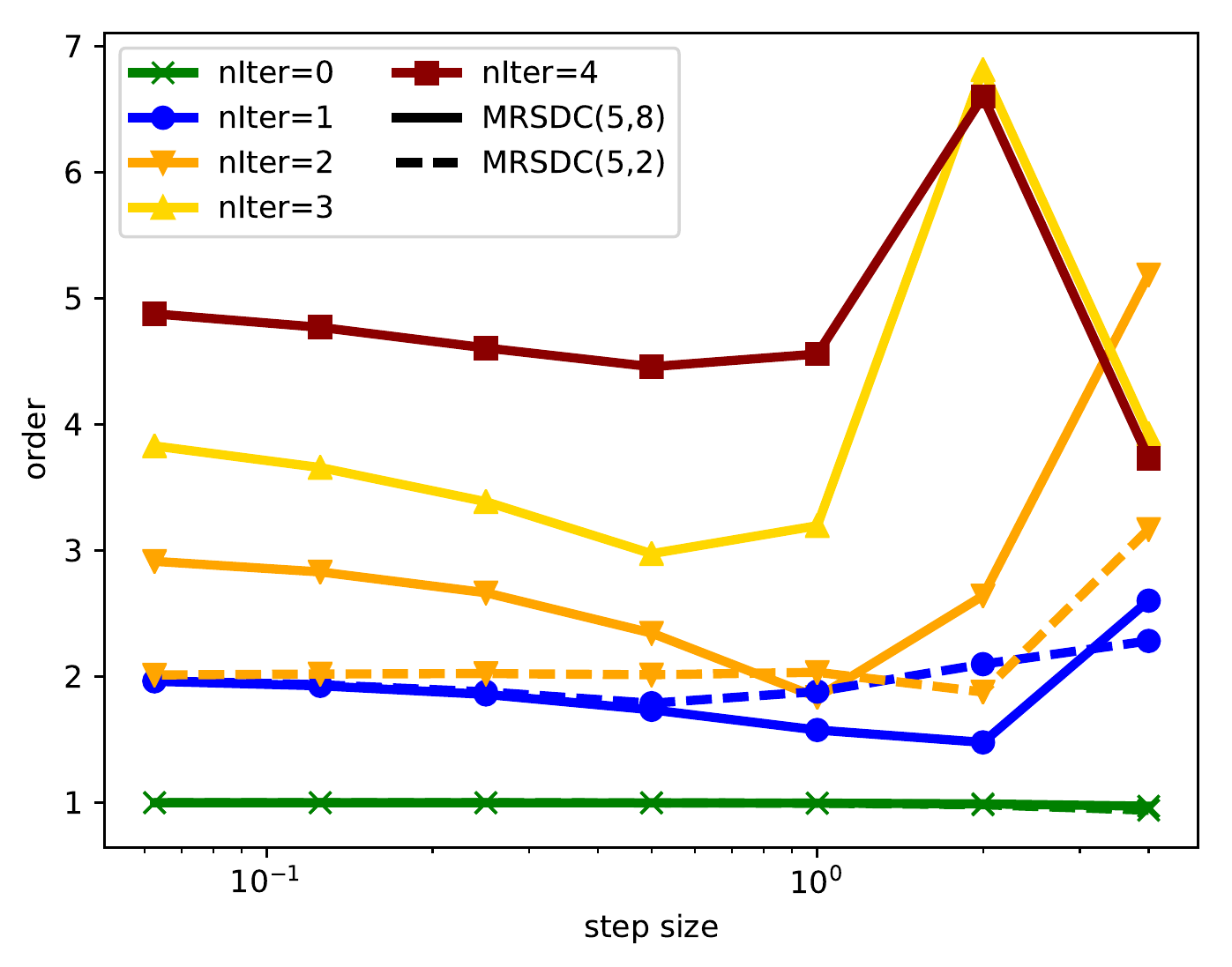}
 \caption{Measured convergence order of MRSDC with $(M,P) = (5,8)$ nodes (solid line) and $(M,P) = (5,2)$ nodes for $k=0,1,2,3,4$ iterations. The order matches the theoretically expected value of $\min(M,P,k+1)$.}
 \label{fig:order_test}
\end{minipage}
\end{figure}

To demonstrate (i), Figure \ref{fig:residual_test} shows the residual~\eqref{eq:residual} plotted against the iteration index $k$ for three different time steps. 
As a guide to the eye, lines proportional to $\Delta t^k$ are shown.
Results from SDC with $M=5$ nodes (dashed lines) and MRSDC with $(M,P) = (5,8)$ nodes (solid lines) are shown.
Residuals are nearly identical for both methods with only small differences for the largest step size $\frac{t_e}{5}$. 
For both methods, the residuals decay proportional to $\Delta t^k$ so that smaller time steps lead to faster convergence.
Eventually, both methods reproduce the collocation solution up to machine precision.

To demonstrate (ii), Figure~\ref{fig:order_test} shows the measured order of convergence for a wide range of time steps for MRSDC with $(M,P) = (5,2)$ nodes (dashed lines) and $(M,P) = (5,8)$ nodes (solid lines). 
Color indicates the number of iterations, ranging from $k=0$ (predictor only) up to $k=4$ (predictor plus four correction sweeps).
The exact solution against which we compare is computed by running the method with a time steps many orders of magnitude smaller. 
Since order is defined for step sizes approaching zero, we see some inconsistent behavior for larger time step sizes. 
With decreasing step size, however, measured order approaches the theoretically expected order of $\min(M, P, k+1)$. 
This illustrates that MRSDC, just as single-rate SDC, improves formal order by one per iteration up to the order of the underlying quadrature rules.

%
%
\section{Real-time simulation of the 3D fully coupled machine}\label{seq:numericalResults}
We now demonstrate that MRSDC implemented in DUNE can accurately solve the 3D fully coupled problem accurately with look-ahead factors $\eta \gg 1$.
Further, we show that the multi-rate time stepping produces a smoother temperature field and more accurate approximations of deformations.
We focus on the error from the time discretization since the FEM approach used in space is standard and its analysis is now textbook material.

The physical parameters and geometries are the same as in Naumann et al.~\cite{Naumann2016}, except for the movement profile of the stock and the simulation time.
We let the stock move according to 
\begin{align}
 s(t)=&a\sin\left(\frac{2\pi}{\varepsilon}t\right) +s_0,
\end{align}
in meters with $a=$\SI{0.495}{\meter} and $s_0$=\SI{0.505}{\meter}.
The movement of the stock is periodic with each period having length $\varepsilon = 24$\SI{}{\second}.
We simulate $10$ periods for a final time of $T=240$\SI{}{\second}.
Our MRSDC time stepping uses $(M,P) = (3,2)$ quadrature nodes and $24$ time steps of length $\Delta t = 1.0$\SI{}{\second} per period.
As initial data we use the stationary temperature profile for the machine at rest but subject to thermal coupling with the environment and cooling equipment
\begin{align}
 \matr{M}\ve{T}(0)=&\ve{f}^I(\ve{T}(0))+\ve{g}(0)\,.
\end{align}

\begin{figure}[t!]
\includegraphics[width=.8\textwidth]{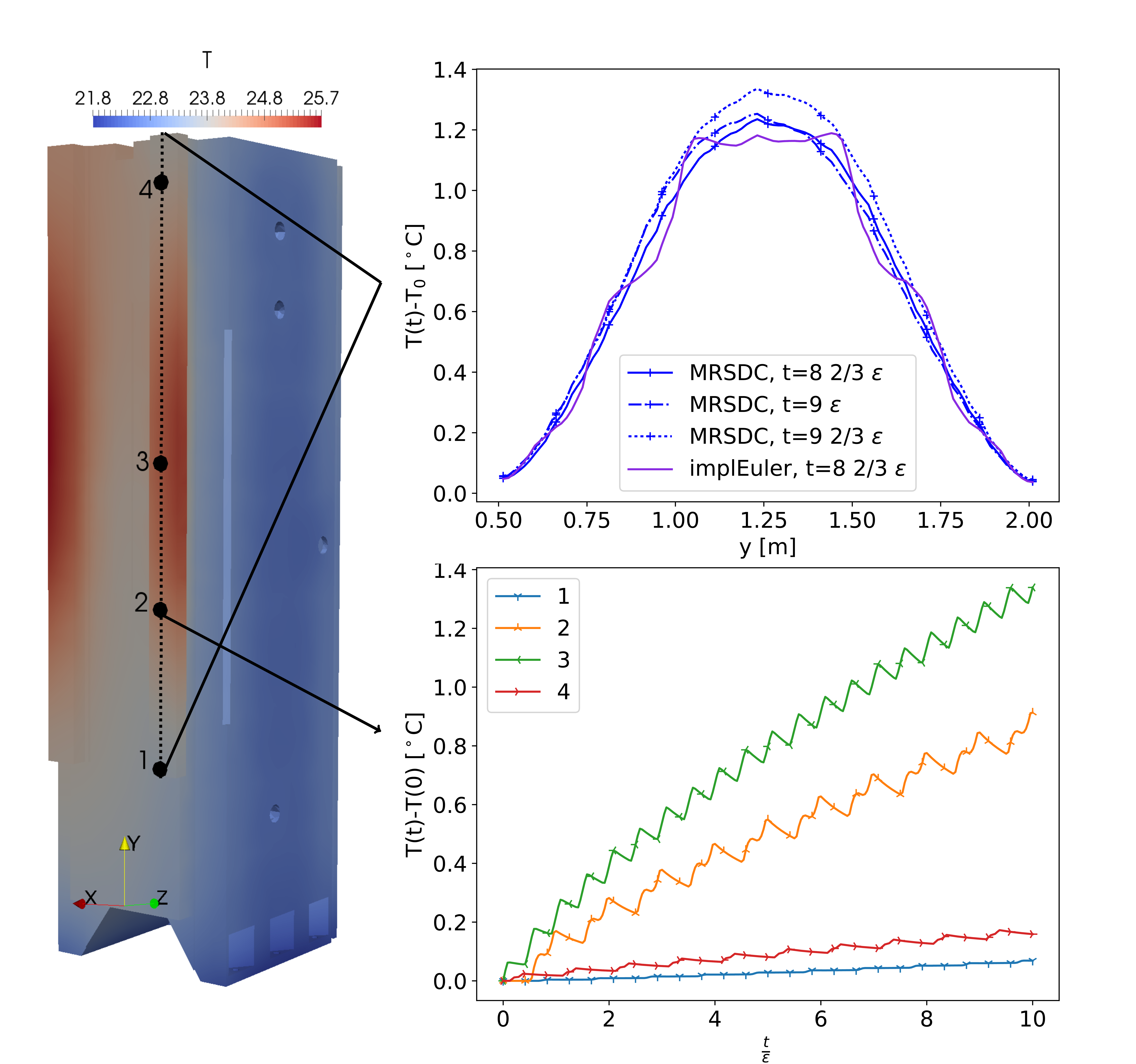}
 \caption{Simulated temperature field on the fixed part of the machine after $t=$\SI{240}{\second} simulated time (left figure). The upper right figure shows the cross-section of the temperature field along the black line on the rail at four different times. Note that at times $t=\frac{2}{3}\varepsilon, \frac{4}{3}\varepsilon, \ldots$, the stock is at the lower end of the rail. At times $t = \varepsilon, 2 \varepsilon, \ldots$ it is at the centre of the rail. The lower right figure shows the temperature over time at the four points indicated by the black dot in the left figure.}
 \label{fig:temperature}  
 \end{figure}
Figure \ref{fig:temperature} shows the temperature field at the end of the simulation at $t = 4$\SI{}{\minute}. 
The maximum temperature (up to \SI{25.7}{\celsius}) is found at the center of rails. 
The fact that we cool the right side of the stand with fixed temperature of \SI{22.0}{\celsius} while the left side is exposed to room temperatures of up to \SI{24.5}{\celsius} creates a slight asymmetry with the left rail being warmer.
Minimum temperatures of \SI{21.8}{\celsius} are found at the floor which has a temperature of \SI{20}{\celsius} and thus removes heat from the machine.
Temperatures at the top of the sides are somewhat higher than towards the bottom because of the small vertical temperature gradient in the environment and the cooling at the bottom.
 
The graphs on the right of Figure~\ref{fig:temperature} illustrate the spatial and temporal variation of the temperature field in specific parts of the machine.
The upper Figure shows the temperature along the indicated cross-section of the right rail at three different times.
For reference, the temperature at $t=8\frac{2}{3}\varepsilon$ computed with implicit Euler is shown as well. 
Toward the center of the rail (at around $y = 1.25$\SI{}{\metre}), the heat generated with each passage of the stock slowly accumulates so that temperature is higher at later times.
After $t=9 \frac{2}{3}$ periods, the temperature has increased by up to \SI{1.4}{\celsius} above the reference temperature.
The strongest warming is seen around the center and the temperature increase becomes less pronounced towards the ends of the rail.
There, the longer time between passages of the stock leaves enough time for the heat to dissipate and only a small increase in temperature of about \SI{0.1}{\celsius} is observed at the upper and lower end.
Furthermore, at full periods, the stock is located at the center of the rail moving upwards, having just heated the lower part, while at two-third periods the stock is near the bottom moving downwards.
This causes a slight shift in the temperature profile at times at full periods, i.e. t=9$\varepsilon$ and $t=10\varepsilon$, compared to times $t = \frac{2}{3}\varepsilon, \frac{4}{3}\varepsilon, \ldots$. 

The lower Figure in~\ref{fig:temperature} shows the transient effects from the moving stock.
Each of the two transits per period (one while moving downwards, one while moving upwards) leads to an increase in temperature, followed by a more gradual decrease due to heat diffusion.
Because the time without transits increases for points away from the center of the rail, there is a longer period of time for the heat to diffuse, leading to less warming.
Transient profiles are therefore not the same throughout the machine but vary with spatial position and models that rely on a separation of spatial and temporal coordinates will not be able to capture this effect.
Points near the center (2 and 3 in Figure~\ref{fig:temperature}) experience a significant net heating of \SI{1.4}{\celsius} and \SI{0.9}{\celsius} respectively over the course of the simulation.
In contrast, points towards the ends of the rails (1 and 4 in Figure~\ref{fig:temperature}) only warm by about \SI{0.1}{\celsius} to \SI{0.2}{\celsius}. 
There is less heating toward the lower end of the rail because of the shorter distance to the cooling floor.

\begin{figure}[t!]
	\centering
	 \includegraphics[width=.8\textwidth]{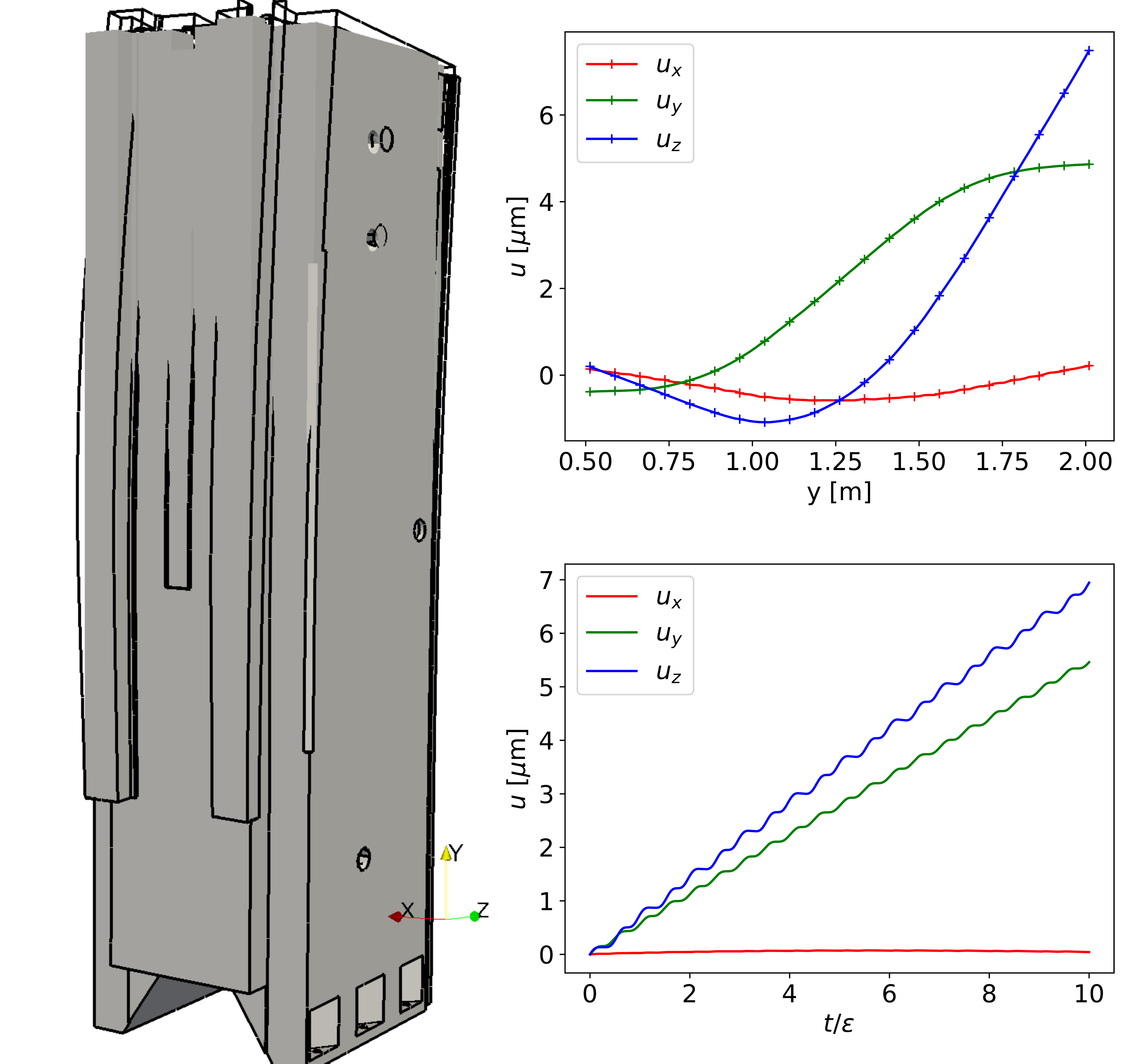}
 \caption{The grey geometry shows the original stand whereas the lines show the deformed stand at time $t=8\frac{2}{3}\varepsilon$ (deformations are exaggerated by a factor of $1e4$ for visibility). Heat is mainly generated at the rail, causing thermal expansion at the front, and therefore the stand bends predominantly in the $z$-direction, away from the rails. The top right figure shows deformations along the same line as in Figure~\ref{fig:temperature} in $x$ (red), $y$ (green) and $z$ (blue) direction. The bottom right figure shows the deformation in point 4 in Figure~\ref{fig:temperature} over time.}
 \label{fig:warpedStand}
\end{figure} 
Deviations from the reference temperature create thermal deformation.
Figure~\ref{fig:warpedStand} shows the deformation of the stand resulting from the temperature field shown in Figure~\ref{fig:temperature}.
Since deformations are of the order of \SI{5}{\micro\metre}, they are exaggerated in the figure by a factor of $10^4$ to make them visible.
The stand mainly bends toward the rear and to the right with stronger deformations at the top.
Because of the transient and inhomogeneous distribution of heat, deformations are not uniform but depend strongly on time and position.
The upper right Figure shows deformations along the cross-section of the right rail indicated in Figure~\ref{fig:temperature}.
The vertical gradient of temperature from to bottom to the top creates significant deformation along the $y$-axis.
Because most of the warming happens at the rails at the front, we observe substantial deformations in $z$ direction.
While relatively small toward the floor, both $z$ and $y$ deformation increase substantially towards the top.
Deformations in $x$ direction due to the slightly asymmetric warming are smaller with a maximum toward the center.

 \begin{figure}[t!]
  \includegraphics[width=\textwidth]{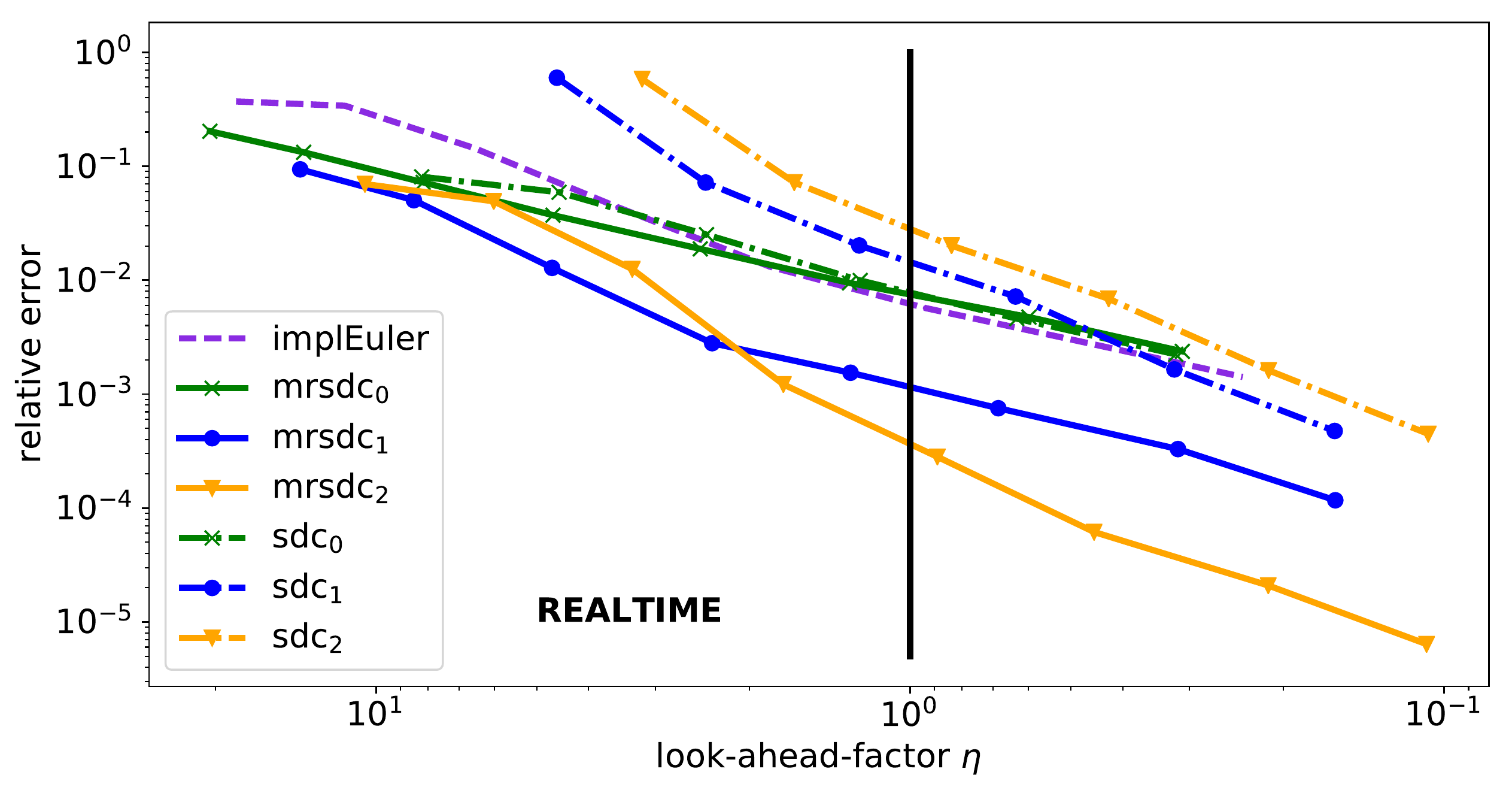}
 \caption{Time-discretisation error for SDC (dash-dotted lines), MRSDC (straight lines) with M=3, P=2 at $t_{end}=10\varepsilon$ and implicit euler (dashed line). The black vertical line indicates $\eta = 1.0$ with simulations on the left running with $\eta > 1$ or faster than real-time.}
 \label{fig:compareMRSDCROS1}
\end{figure}

Next, we analyze performance of MRSDC in terms of work-precision.
Figure~\ref{fig:compareMRSDCROS1} shows achieved time discretization errors (y-axis) in the temperature field versus the look-ahead factor (x-axis) which depends on the wall clock time required to run the simulation at this accuracy. 
Lower errors require better resolution which results in longer simulations and therefore smaller $\eta$.
The threshold between faster than real-time ($\eta > 1$) and slower than real-time ($\eta < 1$) is indicated by a vertical black line.
Three classes of method are investigated: multi-rate SDC (MRSDC, solid lines), single-rate SDC (dash-dotted lines) and implicit Euler (dashed line).
MRSDC uses $(M,P) = (3,2)$ nodes and $k=0, 1, 2$ iterations while SDC uses $M=3$ nodes and also up to two iterations.
Note that the fast component $\matr{M_B}_{\text{fix}}(t)$ from~\eqref{eq:coupledFEM} -- which is treated explicitly with a small step in the multi-rate integrator -- is included in the implicit part in single-rate SDC. 
This means that SDC requires the reassembly of the Jacobian that MRSDC avoids, creating substantial overhead.

For $k=0$, both SDC and MRSDC show first order convergence in the faster and slower than real-time regime.
Both methods deliver about the same efficiency, being slightly better than implicit Euler, but MRSDC has a slight advantage for large values of $\eta$.
For higher order and $k=1,2$, single-rate SDC is substantially less efficient than MRSDC, producing larger errors for the same $\eta$.
Single-rate SDC is also mostly less efficient than implicit Euler except for values $\eta \ll 1$.
MRSDC with $k=1, 2$ is still in the pre-asymptotic regimes for look-ahead factors larger than one, not yet showing the theoretical convergence order.
A clear difference in the slopes of the error lines emerges only for factors of around $\eta \approx 1$ and smaller.
Nevertheless, MRSDC with $k=1,2$ iterations is more efficient than first order MRSDC for most values between $\eta = 10$ to $\eta = 1$ and significantly more efficient than implicit Euler.
Only for very coarse resolutions and values of $\eta > 10$ is there no clear gain from higher order MRSDC with all $k=0,1,2$ showing roughly the same performance.
Still, the multi-rate integration makes MRSDC more efficient than simple implicit Euler, delivering substantially more accurate solutions for the same look-ahead factors.
For slower than real-time simulations with $\eta<1$, MRSDC eventually shows its theoretical order of convergence which significantly widens the performance gap compared to first order implicit Euler.

\begin{figure}[t!]
\includegraphics[width=\textwidth]{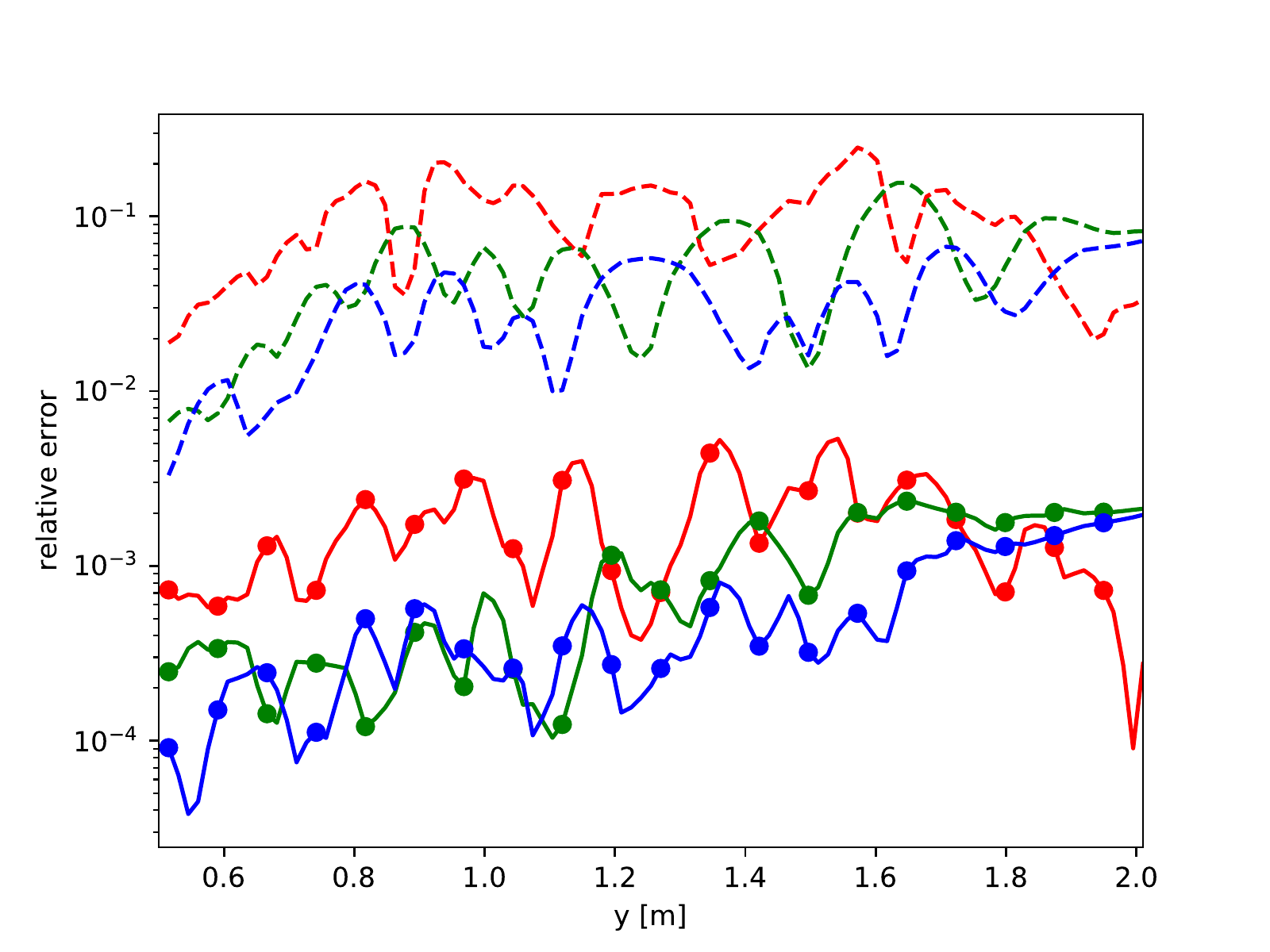}
 \caption{Errors in the displacements in $x$ (red), $y$ (green) and $z$ (blue) direction at time t=$8\frac{2}{3}\varepsilon$ along the line in figure \ref{fig:temperature}. The dislocation correspond to the temperatures in the right top image of Figure \ref{fig:temperature}. Lines with markers show the MRSDC(3,2) solution with one iteration, dashed lines represent the implicit euler. The errors in the displacements from the MRSDC solutions are about two magnitudes smaller. Results are shown for simulations using 12 time steps per period, which corresponds to a look-ahead-factor $\eta\approx 5$ for both methods.}
 \label{fig:dislocationOnLine} 
\end{figure}
We can relate the accuracy of the representation of the temperature field to the accuracy of deformations. 
Figure~\ref{fig:dislocationOnLine} shows the relative error in the deformations in $x$, $y$ and $z$-direction, computed from the gradients of the temperature field, along the cross-section indicated in Figure~\ref{fig:temperature}.
Solid lines marked with circles correspond to MRSDC with $k=1$ iteration and $\eta = 5.0$ while dashed lines indicate implicit Euler with $\eta = 6.5$.
Clearly, the higher accuracy in the computed temperature field translate into significantly more accurate deformations, with errors from MRSDC being at least one order of magnitude smaller than those from implicit Euler.

For the results shown above, we only computed deformations at the end of the simulation. 
In reality, one would have to compute the deformations more frequently.
However, for linear elasticity, computing the deformations reduces to the solution of a large sparse linear system of equations with the temperature profile as right hand side. 
When computing the LU-decomposition of the coefficient matrix at the beginning of the simulation, solving for the deformations requires only a forward-backward solve of lower and upper triangular matrices. 
For the problem studied here, this took about \SI{0.08}{\second} which is negligible compared to the end time \SI{240}{\second} so that more frequent solves will have minimal effect on the reported look-ahead factors.

%
%
\section{Conclusions and outlook}
The paper introduces a multi-rate high-order time stepping method for simulations of heat diffusion in moving machine tools consisting of a fixed stand and a moving stock.
By implementing the algorithm in the open-source FEM framework DUNE, we demonstrate that accurate transient simulations of a FEM model of the fully coupled machine are possible in real-time.
We show that the higher order of multi-rate spectral deferred corrections (MRSDC) improves computational efficiency compared to implicit Euler, even for large time steps where the method does not yet achieve its theoretical order of accuracy.
Time discretization errors of around one percent can be achieved for look-ahead factors of $\eta = 10$.
The results illustrate the potential of solving FEM models fast enough to deliver spatially and temporally resolved temperature fields for online compensation of errors due to thermal deformation.

\paragraph{Outlook}
Open source libraries like the one used in this paper offer the possibility of significant further performance optimization.
Making use of parallelization and high-powered accelerators like graphics processing units or many-core CPUs would require substantial effort but could likely increase look-ahead factor by another order of magnitude or more while maintaining high accuracy.
Exploring novel strategies like parallelization in time~\cite{EmmettMinion2012} could increase $\eta$ even further.
This would eventually allow to use full FEM models as part of filter-based approaches that combine model and measurements into best estimates of the state of a machine~\cite{WegenerEtAl2016}.
Furthermore, coupling a FEM model with a suitable model for error compensation and validating it in a realistic experimental setting~\cite{ThiemEtAl2016} would be an important next step.

\section*{Acknowledgments}
We thankfully acknowledge help from Ansgar Burchardt and Oliver Sander with the DUNE grid-glue library.

\section*{References}

\bibliography{sdc,refs}

\end{document}